\documentclass[aps,pra]{revtex4}
\usepackage{graphicx}

\begin{document}

\title{The 'quantum dialogue' can be eavesdropped under the
intercept-and-resend attack }

\author{ZhongXiao Man \\
{\normalsize Wuhan Institute of Physics and Mathematics, Chinese
Academy of Sciences, Wuhan 430071, China } \\
{\normalsize *Email:zxman@wipm.ac.cn }}

\date{\today}
\maketitle

\begin{minipage}{380pt}
{\bf Abstract}

In this comment we point out that the 'quantum dialogue' (Phys.
Lett. A (in press)) can be eavesdropped under the
intercept-and-resend attack. We also give a revised control mode
to detect this attack. \\


\end{minipage}

Very recently, an entanglement-based protocol for two people to
simultaneously exchange their messages has been proposed by Ba An
Nguyen [1].The original idea has been presented in Ref. [2,3]
earlier. In Ref.[1], Nguyen claims that his protocol is
asymptotically secure against the disturbance attack, the
intercept-and-resend attack and the entangle-and-measure attack.
In this comment, we show this protocol is insecure under the
intercept-and-resend attack and we give a revised protocol to
detect this attack at the same time.

In Nguyen's protocol [1], Bob first produces a large enough number
of Einstein-Podolsky-Rosen (EPR) pairs, all in the state
$|\Psi_{0,
0}\rangle_{ht}=\frac{1}{\sqrt{2}}(|\downarrow\rangle_{h}|\uparrow\rangle_{t}
+|\uparrow\rangle_{h}|\downarrow\rangle_{t}$, where $h$ stands for
"{\it home}", $t$ for "{\it travel}" while $|\downarrow\rangle$
and $|\uparrow\rangle$ characterize two degrees of freedom of a
qubit. Bob encodes his bits $(k_{n}, l_{n})$ ($k_{n}, l_{n}
\in\{0,1\}$) by applying an operation $C_{k_{n},l_{n}}$ on the
state $|\Psi_{0,0}\rangle_{ht}$, keeps one qubit (home qubit) with
him and sends another (travel qubit) to Alice. Then Bob lets Alice
know that. Alice confirms Bob that she received a qubit. Alice
encodes her bits $(i_{n}, j_{n})$ ($i_{n}, j_{n}\in\{0,1\}$) by
performing an operation $C_{i_{n},j_{n}}$ on this travel qubit,
then sends it back to Bob. When Bob receives this encoded travel
qubit he performs a Bell basis measurement on this qubit pair and
waits for Alice to tell him that was a run in message mode (MM) or
in control mode (CM). In MM run, Bob decode Alice's bits and
announces his Bell basis measurement result ($x_{n}, y_{n}$) to
let Alice decode his bits. In CM run, Alice reveals her encoding
value to Bob to check the security of their dialogue.

However this security checking can not detect Eve's
intercept-and-resend attack. Let us suppose Eve prepares a number
of EPR pairs in one of the four Bell states. Then Eve intercepts
the travel qubit which has encoded Bob's bits $(k_{n}, l_{n})$ and
replaces it with a qubit in her prepared EPR pair. She sends this
{\it fake} travel qubit to Alice and retains another with her.
Although Alice confirms Bob that she received a qubit, she can not
distinguish this qubit whether it is the original qubit or the
fake qubit. Alice encodes her bits $(i_{n}, j_{n})$ on this {\it
fake} qubit and sends it back to Bob. Then Eve intercepts this
encoded {\it fake} qubit and performs a Bell basis measurement on
it and the retained qubit. Since Eve knows the state of her
prepared EPR pair, she can conclude Alice's bits $(i_{n}, j_{n})$
and the encoding operation $C_{i_{n},j_{n}}$. And then Eve
performs a same operation as Alice on the original travel qubit
and sends it back to Bob. Consequently, if it is a MM run, after
Bob announces publicly his Bell basis measurement result ($x_{n},
y_{n}$), Eve can deduce Bob's bits: $k_{n}=|x_{n}-i_{n}|,
l_{n}=|y_{n}-j_{n}|$. Hence, in the MM run, Eve eavesdrops
completely the contents of the Bob and Alice's dialogue. Actually,
Eve replaces Alice to perform the encoding operation
$C_{i_{n},j_{n}}$ on the travel qubit , as a result, when Alice
publicly reveals the value $(i_{n}, j_{n})$ for Bob to check
eavesdropping, both $i_{n}=|x_{n}-k_{n}|$ and
$j_{n}=|y_{n}-l_{n}|$ still hold. Accordingly, in the CM run the
legitimate users, Alice and Bob, can not detect this
intercept-and-resend eavesdropping.

Now we give a revised control mode (RCM) to detect this
intercept-and-resend eavesdropping. At first, we define the four
Bell states as follows.
\begin{eqnarray}
|\Psi_{0,0}\rangle =\frac{1}{\sqrt{2}}(|0\rangle |1\rangle
+|1\rangle |0\rangle )=\frac{1}{\sqrt{2}}(|+\rangle |+\rangle
-|-\rangle |-\rangle ),\\
|\Psi_{0,1}\rangle=\frac{1}{\sqrt{2}}(|0\rangle |1\rangle
-|1\rangle |0\rangle )=\frac{1}{\sqrt{2}}(|+\rangle |-\rangle
-|-\rangle |+\rangle ),\\
|\Psi_{1,0}\rangle =\frac{1}{\sqrt{2}}(|0\rangle |0\rangle
+|1\rangle |1\rangle =\frac{1}{\sqrt{2}}(|+\rangle |+\rangle
+|-\rangle |-\rangle ),\\
|\Psi_{1,1}\rangle =\frac{1}{\sqrt{2}}(|0\rangle |0\rangle
-|1\rangle |1\rangle )=\frac{1}{\sqrt{2}}(|+\rangle |-\rangle
+|-\rangle |+\rangle ),
\end{eqnarray}
where$|+\rangle=\frac{1}{\sqrt{2}}(|0\rangle+|1\rangle),
|-\rangle=\frac{1}{\sqrt{2}}(|0\rangle-|1\rangle).$

At the beginning, Bob prepares some EPR pairs in the state
$|\Psi_{0,0}\rangle$. Suppose that

Alice's message=\{$(i_{1},j_{1}), (i_{2},j_{2},..., (i_{N},
  j_{N})\}$

Bob's message=\{$(k_{1},l_{1}), (k_{2},l_{2},..., (k_{N},
l_{N})$\}

Bob encodes his bits $(k_{n},l_{n})$ on the EPR pair
$|\Psi_{0,0}\rangle$ by performing an operation $C_{k_{n},l_{n}}$
on one qubit. Then Bob sends one qubit to Alice (travel qubit) and
keeps another with him (home qubit). Alice confirms Bob that she
received a qubit. In the RCM run, Alice and Bob check
eavesdropping by following procedure: (a)Alice chooses randomly
one of the two sets of measuring basis (MB), i.e.,
$\ss_{z}=\{{|0\rangle, |1\rangle}\}$ and $\ss_{x}=\{|+\rangle,
|-\rangle\}$, to measure the travel qubit. (b) Alice tells Bob
which MB she has chosen and the outcomes of her measurement. (c)
Bob uses the same measuring basis as Alice to measure the home
qubit and checks with the results of Alice. If no eavesdropping
exists, their results should be correlated according to Eq.(1-4).
As an example, if $k_{n}=0$ and $l_{n}=1$, Bob applies the
operation $C_{0,1}$ on one qubit in the EPR pair
$|\Psi_{0,0}\rangle$, then the state $|\Psi_{0,0}\rangle$ is
changed into $|\Psi_{0,1}\rangle$. If there is no Eve in line the
results must be completely opposite, i.e., if Bob obtains
$|0\rangle$ ($|+\rangle$), then Alice gets $|1\rangle$
($|-\rangle$). By this revised control mode, any
intercept-and-resend eavesdropping can be detected.

 This work is supported by the National Natural Science
Foundation of China under Grant Nos. 10304022 and 10374119.

{\bf References}

\noindent[1] B. A. Nguyen, Phys. Lett. A (in press).

\noindent[2] Z. J. Zhang quant-ph/0403186 (2004).

\noindent[3] Z. J. Zhang and Z. X. Man quant-ph/0403215 (2004).

\enddocument